%% file: arxiv.tex
\PassOptionsToPackage{dvipsnames}{xcolor}
\documentclass[10pt,sigconf,nonacm,screen]{acmart}

\usepackage[all]{nowidow}
\usepackage{xcolor}
\usepackage{subcaption}
\usepackage{hyperref}
\usepackage{acronym}
\usepackage{tikz}

\usepackage[capitalise,noabbrev]{cleveref}
\crefformat{section}{\S#2#1#3} 
\crefformat{subsection}{\S#2#1#3}
\crefformat{subsubsection}{\S#2#1#3}
\crefrangeformat{section}{\S#3#1#4 to~\S#5#2#6}
\crefrangeformat{subsection}{\S#3#1#4 to~\S#5#2#6}
\crefrangeformat{subsubsection}{\S#3#1#4 to~\S#5#2#6}

\newcommand*\circled[1]{\tikz[baseline=(char.base)]{
\node[shape=circle,draw,inner sep=0.75pt,scale=0.8,font=\fontfamily{phv}\selectfont] (char) {#1};}%
}

\begin{document}

\author{Felix Moebius}
\affiliation{%
    \institution{TU Berlin \& ECDF}
    \city{Berlin}
    \country{Germany}}
\email{fmo@mcc.tu-berlin.de}

\author{Tobias Pfandzelter}
\affiliation{%
    \institution{TU Berlin \& ECDF}
    \city{Berlin}
    \country{Germany}}
\email{tp@mcc.tu-berlin.de}

\author{David Bermbach}
\affiliation{%
    \institution{TU Berlin \& ECDF}
    \city{Berlin}
    \country{Germany}}
\email{db@mcc.tu-berlin.de}

\title{Are Unikernels Ready for Serverless on the Edge?}

\begin{abstract}
    Function-as-a-Service (FaaS) is a promising edge computing execution model but requires secure sandboxing mechanisms to isolate workloads from multiple tenants on constrained infrastructure.
    Although Docker containers are lightweight and popular in open-source FaaS platforms, they are generally considered insufficient for executing untrusted code and providing sandbox isolation.
    Commercial cloud FaaS platforms thus rely on Linux microVMs or hardened container runtimes, which are secure but come with a higher resource footprint.

    Unikernels combine application code and limited operating system primitives into a single purpose appliance, reducing the footprint of an application and its sandbox while providing full Linux compatibility.
    In this paper, we study the suitability of unikernels as an edge FaaS execution environment using the Nanos and OSv unikernel tool chains.
    We compare performance along several metrics such as cold start overhead and idle footprint against sandboxes such as Firecracker Linux microVMs, Docker containers, and secure gVisor containers.
    We find that unikernels exhibit desirable cold start performance, yet lag behind Linux microVMs in stability.
    Nevertheless, we show that unikernels are a promising candidate for further research on Linux-compatible FaaS isolation.
\end{abstract}

\maketitle

\input{sections/1_introduction.tex}
\input{sections/2_background.tex}
\input{sections/3_approach.tex}
\input{sections/4_results.tex}
\input{sections/5_discussion.tex}
\input{sections/6_relwork.tex}
\input{sections/7_conclusions.tex}

\begin{acks}
    Funded by the \grantsponsor{BMBF}{Bundesministerium für Bildung und Forschung (BMBF, German Federal Ministry of Education and Research)}{https://www.bmbf.de/bmbf/en} -- \grantnum{BMBF}{16KISK183}.
\end{acks}

\balance

\bibliographystyle{ACM-Reference-Format}
\bibliography{bibliography.bib}

\end{document}

%% file: sections/1_introduction.tex
\section{Introduction}
\label{sec:introduction}

Serverless computing with Functions-as-a-Service (FaaS) allows developers to deploy scalable applications as small, stateless functions that are invoked in a sandbox environment based on incoming events, with resource management and scaling handled by the underlying FaaS platform~\cite{baldini_2017,castro_2019}.
On-demand resource allocation and per-invocation isolation make it an ideal fit for multi-tenant edge environments, where limited resources must be shared between services~\cite{paper_bermbach2017_fog_vision,rausch2019towards,paper_pfandzelter2020_tinyfaas,paper_bermbach2021_auctionwhisk}.

Docker containers are commonly used as a FaaS sandbox mechanism as they are lightweight and compatible with most applications~\cite{palade2019evaluation,paper_pfandzelter2020_tinyfaas}.
However, they lack the security and isolation controls necessary when running services from multiple tenants on a single host, making them unsuitable for edge environments~\cite{firecracker_2020,gvisor2023}.
Common approaches in cloud FaaS include deploying small microVMs~\cite{firecracker_2020} or hardened containers that intercept system calls~\cite{gvisor,gvisor-gcf}, yet these sandboxes can introduce considerable invocation overheads and memory footprints that make them unsuitable for the edge~\cite{gvisor-performance}.

An isolation mechanism that has yet to be explored for edge FaaS is the unikernel, which combines application code and limited operating system primitives into a single purpose appliance~\cite{mad_2013_1,mad_2013_2,mad_2014,kuenzer_unleashing_2019,hale_towards_2019,lefeuvre_2021}.
Reducing the footprint of kernel functionality required to run the application service results in a minimal footprint that can boot more quickly while also using less resources~\cite{kuenzer_unikernels_2017_1,kuenzer_unikernels_2017_2,faas-unikernel-practical}.
Nevertheless, unikernels are not yet widely adopted and lack the maturity of container and microVM technology~\cite{cantrill-rant}.

In this paper, we study the suitability of unikernels as edge FaaS execution environments using the Nanos~\cite{nanos} and OSv~\cite{osv_2014} unikernel tool chains.
To the best of our knowledge, this is the first performance study of general-purpose, Linux-compatible unikernels as edge FaaS isolation mechanisms.
Further, we compare performance along multiple relevant metrics against Firecracker Linux microVMs, Docker containers, and the gVisor hardened container runtime.

%% file: sections/2_background.tex
\section{Background}
\label{sec:background}

To understand to what extent unikernels can be used for isolating FaaS workloads on the edge, we first explain the concept of edge FaaS, isolation mechanisms commonly used today, and the architecture of unikernels.

\subsection{FaaS on the Edge}
\label{sec:background:edgefaas}

FaaS applications are composed of small, stateless functions invoked by external events, such as HTTP requests, messages in a communication layer, or changes in an external state management service, e.g., a database~\cite{baldini_2017,castro_2019,jonas2019cloud}.
The small footprint of FaaS functions and the need to externalize their state allows an underlying FaaS platform to rapidly scale-out function instances along with the frequency of incoming events, thus, supporting both elasticity and scale-to-zero.
A key innovation of the FaaS programming model in the context of cloud computing is the pay-per-use billing model where tenants pay only for each millisecond of function activity.
In the context of edge computing, where limited compute and storage resources on the edge of the network host application services with low access latency from clients, FaaS can be used to provision resources in a finely-grained manner while still providing isolation between tenant services~\cite{paper_pfandzelter2020_tinyfaas}.

\subsection{FaaS Isolation}
\label{sec:background:isolation}

\begin{figure*}
    \centering
    \begin{subfigure}{0.48\textwidth}
        \centering
        \includegraphics[width=0.975\columnwidth]{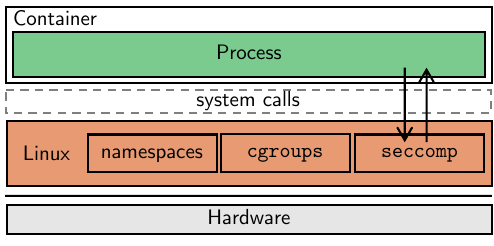}
        \caption{\emph{Containers} isolate processes with namespaces, control groups, and the secure computing facilities.}
        \label{fig:isolation:container}
    \end{subfigure}%
    \hfill
    \begin{subfigure}{0.48\textwidth}
        \centering
        \includegraphics[width=0.975\columnwidth]{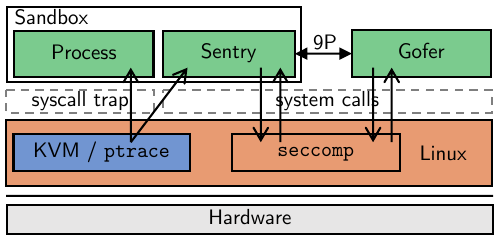}
        \caption{\emph{gVisor} intercepts container system calls and file system access with the Sentry and Gofer components.}
        \label{fig:isolation:gvisor}
    \end{subfigure}%
    \vfill
    \begin{subfigure}{0.48\textwidth}
        \centering
        \includegraphics[width=0.975\columnwidth]{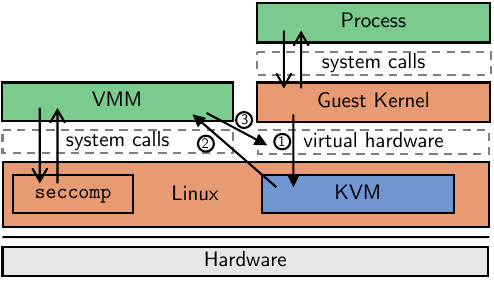}
        \caption{\emph{KVM} virtualization runs an entire guest kernel along with the process. VMMs emulate device access, optionally in a \texttt{seccomp} sandbox.}
        \label{fig:isolation:kvm}
    \end{subfigure}%
    \hfill
    \begin{subfigure}{0.48\textwidth}
        \centering
        \includegraphics[width=0.975\columnwidth]{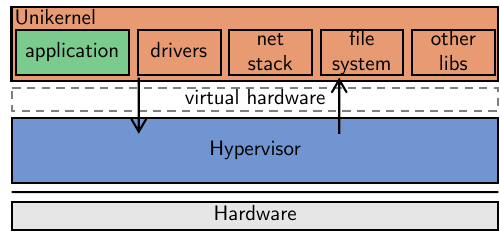}
        \caption{\emph{Unikernels} package application code with virtual drivers or operating system primitives into a bespoke binary that can run directly on a hypervisor.}
        \label{fig:isolation:unikernel}
    \end{subfigure}%
    \caption{Application Service Isolation Mechanisms}
    \label{fig:isolation}
\end{figure*}

Providing isolation between tenants and managing the function instance lifecycle from rapid instantiation to tear-down requires application sandboxing.
A number of different sandboxing mechanisms are usually used for this:
containers, hardened containers, and Linux microVMs.

\subsubsection{Containers}
\label{sec:background:isolation:container}

Open-source FaaS platforms, e.g., OpenWhisk~\cite{openwhisk} or tinyFaaS~\cite{paper_pfandzelter2020_tinyfaas}, rely on containers for isolation, usually also supported by a container orchestration engines such as Kubernetes.
Container tooling such as \texttt{runc} (which is the default runtime for Docker containers) combines Linux facilities such as control groups (\texttt{cgroups}), namespaces, and secure computing (\texttt{seccomp}) facilities into dedicated application execution environments, giving the isolated process the illusion of running on a dedicated system with its own file system and network stack (\cref{fig:isolation:container})~\cite{container_security}.
Control groups restrict processes in their access to resources such as CPU time, disk and network I/O, or memory usage, while secure computing restricts the system calls that a process and its descendants are allowed to make.
Further, namespaces virtualize various shared system resources to provide processes a distinct set of resources without being able to see or access resources from other namespaces.

\subsubsection{Hardened Containers}
\label{sec:background:isolation:gvisor}

While containers are well suited to isolate trusted workloads, they lack the guarantees required to isolate untrusted payloads.
For example, while \texttt{seccomp} can limit the overall attack surface, all containers on a host share the underlying host kernel and are thus susceptible to kernel vulnerabilities~\cite{xiao_2015,leaks_2017,docker-security,bincomp_2019}.
The lack of strong isolation guarantees has led to developments such as Google's gVisor container sandbox, which intercepts all system calls of the processes in a container and forwards them to a user space kernel that emulates the Linux system call interface~\cite{gvisor,gvisor-gcf}.
We refer to this as a \emph{hardened container runtime}, as it limits the interaction between host kernel and container.

As shown in \cref{fig:isolation:gvisor}, the gVisor sandbox adds the \textit{Sentry} process that intercepts and emulates container system calls in user space and the \textit{Gofer} component that provides secured file system access, both implemented through either the \texttt{ptrace} or Linux KVM facilities.
The result is the \texttt{runc}-compatible \texttt{runsc} container runtime.
However, this additional layer of isolation comes at a cost:
gVisor containers incur high overheads when making system calls and degraded I/O performance compared to native execution~\cite{gvisor-performance}.

\subsubsection{microVMs}
\label{sec:background:isolation:microvm}

Hardware virtualization provides greater isolation capabilities than process-level isolation with containers and is the de facto standard for providing isolated compute infrastructure in multi-tenant clouds~\cite{ibm_2018}.
The kernel virtual machine (KVM) shown in \cref{fig:isolation:kvm}, for example, handles the virtualization of CPUs, memory, and fundamental platform devices such as interrupt controllers, but leaves the remaining device emulation to a dedicated user space process commonly referred to as virtual machine monitor (VMM)~\cite{kivity_2007}.
Device accesses are trapped by KVM \circled{1}, the virtual CPU (vCPU) is stopped and control transferred to the VMM process \circled{2}, which will perform the device emulation in software, update the device state as seen by the guest \circled{3}, and hand control back to KVM, which in turn resumes the execution of the vCPU.

Despite this high level of isolation, the increased memory footprint and startup times of virtualization makes it unsuitable for direct use as a FaaS sandbox.
The Firecracker VMM developed for AWS Lambda aims to combine the strong isolation and security guarantees of virtualization with the fast initialization times normally associated with containers~\cite{firecracker_2020}.
The Firecracker device model is limited to \texttt{virtio-block} and \texttt{virtio-net} devices using the \texttt{virtio} MMIO transport instead of the more common PCI variant, which further simplifies its implementation and speeds up the boot process~\cite{russel_2008}.
Firecracker does not require a BIOS or bootloader, and instead implements the architecture-specific Linux boot protocols to directly boot uncompressed guest kernels.
Firecracker can start the VMM process in dedicated mount and network namespaces in a \texttt{seccomp} sandbox with a more limited profile than required for containers, since the requirements for the VMM are limited and known beforehand.

\subsection{Unikernels}
\label{sec:background:unikernel}

When the FaaS workload is a short-lived single process and does not require the majority of features a Linux kernel provides, running an entire guest OS with hardware virtualization will waste considerable host resources.
Unikernels are single-purpose machine images that are specialized for executing one particular application directly on top of a hypervisor, as shown in \cref{fig:isolation:unikernel}~\cite{kuenzer_unleashing_2019,hale_towards_2019,lefeuvre_2021}.
To achieve this, they package the application together with the code required to drive the virtual hardware and provide operating system primitives such as schedulers, memory management, network stacks, or file systems.
Unikernels generally produce lean machine images that can boot significantly faster and have a smaller resource footprint than Linux while still providing the isolation of virtualization, making them an ideal fit for FaaS sandboxes.

Some unikernels provide (near) POSIX-compatibility to applications regardless of the programming language used to implement them\footnote{Note that language-based unikernels, which are closer to \emph{library operating systems} by compiling OS abstractions and execution environment directly alongside application code, are outside the scope of this work as they require modification of the application source, breaking the FaaS abstraction.}~\cite{kuenzer_unleashing_2019}:
The Nanos unikernel aims at full binary compatibility with Linux by implementing the Linux system call interface and providing an ELF loader to load arbitrary Linux executables at runtime~\cite{nanos}.
The OSv unikernel provides a dynamic linker to link an application to custom implementations of core system libraries such as  \texttt{libc} or \texttt{libpthread} at runtime as many programming languages do not make system calls directly, and instead link applications against the platform's \texttt{libc} and use its wrapper functions~\cite{osv_2014}.

%% file: sections/3_approach.tex
\section{Approach and Setup}
\label{sec:approach}

\begin{figure}
    \centering
    \includegraphics[width=0.99\columnwidth]{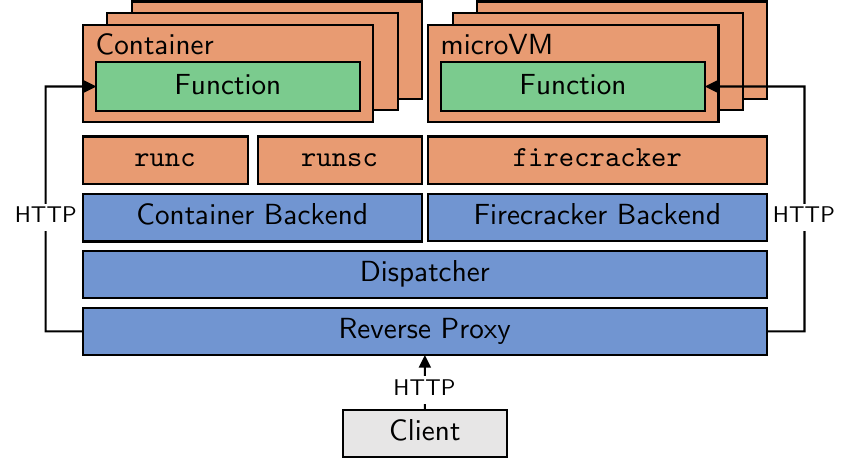}
    \caption{The FaaS sandbox experiment harness can route requests to \texttt{runc}, \texttt{runsc}, or Linux and unikernel microVM workers that are created by a dispatching component.}
    \label{fig:harness}
\end{figure}

To evaluate the viability of using unikernels in FaaS edge deployments, we implemented a single-node FaaS system experiment harness that can instantiate function execution environments based on Linux microVMs, Nanos and OSv unikernels, as well as Linux containers using both \texttt{runc} and \texttt{runsc} container runtimes.
Our experiment harness follows the design of the lightweight single-node edge FaaS platform tinyFaaS~\cite{paper_pfandzelter2020_tinyfaas}, as shown in \cref{fig:harness}:
Clients invoke functions over HTTP against the reverse proxy, which invokes the actual sandboxed function over HTTP, first instructing the dispatcher to create a new function instance if none is available.
This dispatcher can be configured with a container or microVM backend:
The container backend uses either \texttt{runc} or \texttt{runsc} to start a new function instance in a container, whereas the microVM backend instructs the Firecracker VMM to start a new Linux or unikernel (Nanos or OSv) microVM for the function.
In either case, we assume that the container image, Linux root file system, or unikernel image is already configured and available.
Preliminary experiments revealed that this FaaS system adds a mean 0.9ms invocation overhead to function calls for fully receiving the incoming request before forwarding it, determining if a function instance is available to serve the request, and establishing an additional TCP connection to that function instance.

Container and Linux microVM images are based on the minimal \emph{alpine Linux} base image and Linux microVMs use a minimal Linux kernel in v5.10 adapted from the recommended Firecracker configuration.
We configure gVisor to use the KVM backend, which offers better performance than \texttt{ptrace}~\cite{gvisor-production}.
We use a slightly modified version of the Nanos unikernel\footnote{\url{https://github.com/felixmoebius/nanos/tree/faas}} that allows us to provide network configuration and callback URLs from the kernel command line, and we use the \texttt{ops} tool to construct and deploy Nanos unikernels.
Finally, we use the lightweight read-only file system implementation of OSv as we deem this sufficient for most FaaS workloads.
Our experiment harness is written in Rust and available as open-source software along with all other evaluation artifacts.\footnote{\url{https://github.com/OpenFogStack/unikernel-edge-faas}}

We use this experiment harness to run a number of experiments, using the different backend options.
As hardware platform, we use an 11th generation Intel NUC with an 8-core Intel i5-1135G7 processor and 64GiB of memory, running Ubuntu 23.04 with a v6.2 kernel.
We make all requests to the FaaS system locally on the same machine, thus removing any overhead incurred by a physical network connection.
All functions in our evaluation are limited to a single CPU and 512 MiB of memory.

%% file: sections/4_results.tex
\begin{figure*}
    \centering
    \begin{subfigure}{0.49\textwidth}
        \centering
        \includegraphics[width=0.975\columnwidth]{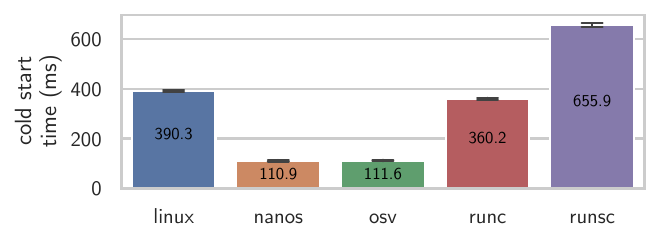}
        \caption{Go no-op function}
        \label{fig:cold-start-single:go}
    \end{subfigure}%
    \hfill
    \begin{subfigure}{0.49\textwidth}
        \centering
        \includegraphics[width=0.975\columnwidth]{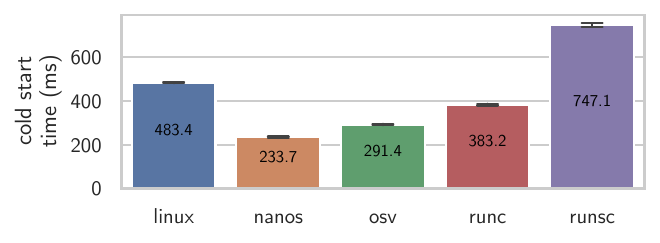}
        \caption{Node.js no-op function}
        \label{fig:cold-start-single:node}
    \end{subfigure}%
    \caption{Mean time for single cold starts in different execution environments (whiskers show 95\textsuperscript{th} percentile confidence interval)}
    \label{fig:cold-start-single}
\end{figure*}

\section{Experiments and Findings}
\label{sec:results}

Using synthetic benchmarks for isolated measurements against our sandboxing mechanisms, we investigate single and burst function cold start latency, resource requirements for sandbox initiation, idle resource usage, CPU and memory performance, as well as network I/O and file system read performance.
To this end, we use FaaS functions written in Go (statically compiled) and Node.js (JIT compilation), two popular options in commercial FaaS offerings~\cite{paper_pfandzelter2022_streamingfunctions}.
We repeat each experiment 100 times to ensure robustness.

\subsection{Cold Start Latency}
\label{sec:results:cold-single}

To measure cold start latency, we deploy simple Go and Node.js no-op functions that answer requests immediately without performing additional computation.
We invoke these functions from a client that measured request completion time, thus essentially measuring the overhead of sandbox instantiation and starting the function executable.
As shown in \cref{fig:cold-start-single}, both Nanos and OSv are able to reduce the cold start latency to around 110ms with the Go function, which is less than a third of the time it takes to boot the microVM and start the function handler with Linux.
Docker with the default \texttt{runc} runtime is slightly faster than the Linux microVM, while the gVisor version is the slowest at around 656ms.

The differences are less pronounced for the Node.js function, yet OSv and Nanos still take more than twice as long as the Linux microVM.
\texttt{runc} containers have direct access to the host page cache, which already contains the large Node.js executable when repeatedly starting the same container image, partially explaining the comparatively small overhead.
We do not observe these benefits with gVisor-based containers.

These results demonstrate clear advantages for unikernels in reducing the sandbox startup latency.
We also find that the time required for starting a container with \texttt{runc} is unexpectedly large compared to unikernels.
Note that for Linux microVMs, cold start latency can be mitigated by using pre-booted VMs, which we further discuss in \cref{sec:discussion:optimization}.

\subsection{Burst Cold Starts}
\label{sec:results:cold-burst}

\begin{figure*}
    \centering
    \begin{subfigure}{0.49\textwidth}
        \centering
        \includegraphics[width=0.975\columnwidth]{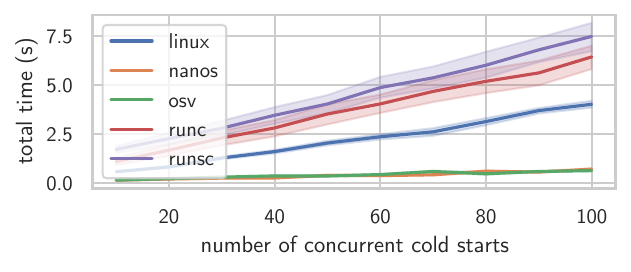}
        \caption{Go no-op function}
        \label{fig:cold-start-burst:go}
    \end{subfigure}%
    \hfill
    \begin{subfigure}{0.49\textwidth}
        \centering
        \includegraphics[width=0.975\columnwidth]{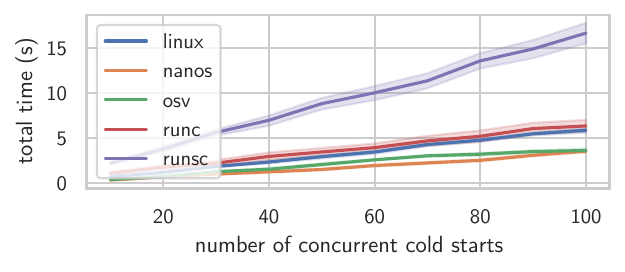}
        \caption{Node.js no-op function}
        \label{fig:cold-start-burst:node}
    \end{subfigure}%
    \caption{Mean cold start time for $n$ concurrent starts}
    \label{fig:cold-start-burst}
\end{figure*}

To investigate how well the different sandbox mechanisms scale when handling cold starts from multiple concurrent function invocations, we perform them in batches of increasing sizes.
We issue 10 to 100 concurrent invocations of the same no-op functions, triggering a cold start for each invocation (we do not reuse function instances).

As shown in \cref{fig:cold-start-burst}, Nanos and OSv show better scaling behavior than all other execution environments.
When starting 100 instances of the Go function at the same time, all requests finish in less than one second on Nanos and OSv, with a mean of around 700ms on Nanos and 630ms for OSv.
With the Linux microVM, starting 100 instances take around 4.3s on average.
Docker with \texttt{runc} and \texttt{runsc} takes longer to start 100 instances with around 6.5s and 7.5s respectively.

The differences between the Linux microVM and the unikernel environments are again less pronounced with the Node.js function.
Here, Nanos and OSv need a mean 3.6s to start 100 Node.js function instances, which is still almost twice as fast as the Linux microVM (5.9s).
Docker with \texttt{runc} takes around 6.4s on average, whereas the \texttt{runsc} runtime performed significantly worse (16.7s).

Overall, these results show that the light-weight nature and the fast boot times of unikernels let them handle multiple concurrent cold starts significantly better than Linux microVMs and gVisor-based containers.

\subsection{Sandbox Initiation Resource Usage}
\label{sec:results:instantiation}

\begin{figure*}
    \centering
    \begin{subfigure}{0.49\textwidth}
        \centering
        \includegraphics[width=0.975\columnwidth]{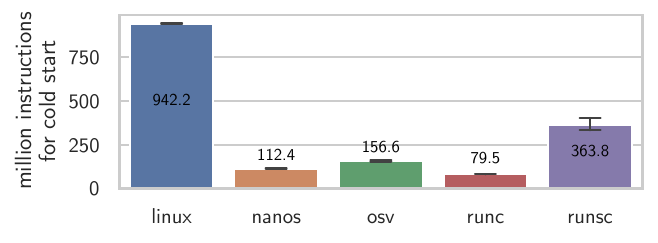}
        \caption{Go no-op function}
        \label{fig:cold-start-instructions:go}
    \end{subfigure}%
    \hfill
    \begin{subfigure}{0.49\textwidth}
        \centering
        \includegraphics[width=0.975\columnwidth]{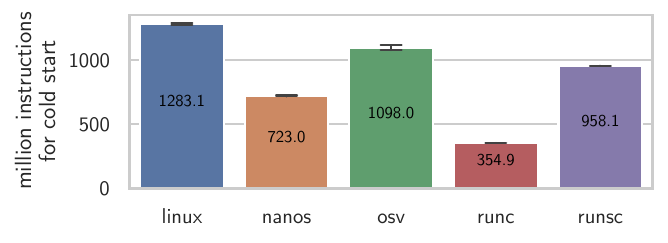}
        \caption{Node.js no-op function}
        \label{fig:cold-start-instructions:node}
    \end{subfigure}%
    \caption{Millions of instructions executed to instantiate a function instance in different execution environments}
    \label{fig:cold-start-instructions}
\end{figure*}

In addition to being short, cold starts should also be resource-efficient on the resource-constrained edge, occupying the CPU as little as possible.
To quantify the resource strain of cold starts, we use \texttt{perf}\footnote{\url{https://perf.wiki.kernel.org/}} to record the number of instructions executed for cold starts with different execution environments.
For Linux, Nanos, and OSv microVMs executing on Firecracker, we record the instructions executed by the Firecracker process, including the guest VM, using a custom \texttt{bpftrace}\footnote{\url{https://github.com/iovisor/bpftrace}} script to probe when a function instance is ready to receive an invocation.
For similar measurements with containers, we directly invoke and trace the \texttt{runc} and \texttt{runsc} container runtimes without using Docker.
As \texttt{perf} tracing includes child processes, tracing \texttt{runsc} also includes the Sentry and Gofer components.
We fully evict all relevant files from the host page cache to reduce the impact of page caching, yet find that this did not lead to a significant slowdown.

As shown in \cref{fig:cold-start-instructions}, instantiating the Go function on Linux microVMs takes considerably more instructions than on both unikernels, with 8.5 and 6 times as many instructions executed compared to Nanos and OSv, respectively.
Although we measure 30\% slower cold start times for gVisor compared to Linux microVMs, gVisor requires 61\% \emph{fewer} instructions than Linux to instantiate the Go function.

Unsurprisingly, we find that the Node.js runtime adds considerable overhead.
Importantly, this overhead differs widely between execution environments.
We also observe almost twice as many additional instructions between the Go and Node.js functions on gVisor compared to the Linux microVM, even though they use the exact same versions of Go and Node.js.
This indicates that starting a complex language runtime such as Node.js causes varying overhead in different execution environments, which is likely a result of how they implement the required operating system functionality.
Linux generally appears to handle this task better than the other systems, as both \texttt{runc} containers and the Linux microVM show a smaller difference between the Go and Node.js functions.

Our results show that unikernels can significantly reduce the cost of cold starts in terms of CPU usage, by up to 8.5 times compared to a Linux microVM and up to 2.5 times compared to gVisor.
However, we also find that when using a heavy-weight function application, such as one based on Node.js, the advantage is less pronounced as the overhead of starting the function application can be multiple times larger than that of setting up the sandboxed execution environment.

\subsection{Idle Resource Usage}
\label{sec:results:idle}

Although FaaS enables scaling to zero, FaaS platforms typically keep `warm' function instances available to serve subsequent requests more quickly~\cite{firecracker_2020}.
Ideally, these warm instances should have a low CPU and memory footprint in order to~(a)~not degrade the performance of co-located function execution in the resource constrained environment and (b)~enable FaaS platforms to keep as many warm instances available as possible without running into resource contention.

\begin{figure*}
    \centering
    \begin{subfigure}{0.49\textwidth}
        \centering
        \includegraphics[width=0.975\columnwidth]{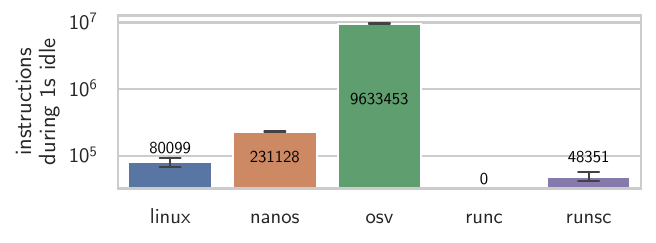}
        \caption{Go no-op function}
        \label{fig:idle-instructions:go}
    \end{subfigure}%
    \hfill
    \begin{subfigure}{0.49\textwidth}
        \centering
        \includegraphics[width=0.975\columnwidth]{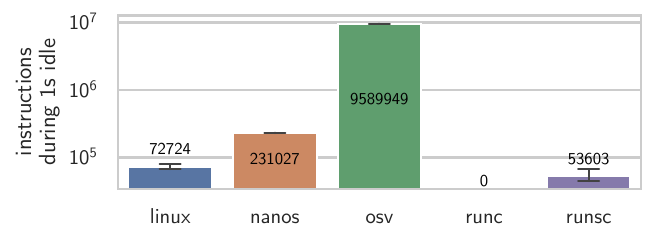}
        \caption{Node.js no-op function}
        \label{fig:idle-instructions:node}
    \end{subfigure}%
    \caption{Instructions executed during a 1000ms idle period (log scale)}
    \label{fig:idle-instructions}
\end{figure*}

\subsubsection*{CPU}
In theory, none of the sandboxes we evaluate should be doing any meaningful work while the application function is idle.
To assess CPU usage, we use \texttt{perf} to record the number of instructions executed within a one-second interval 15 seconds after starting the no-op function instance.
This includes VMM instructions for Firecracker or Sentry and Gofer processes for gVisor.
The results in \cref{fig:idle-instructions} show the lowest instruction count for gVisor (50k executed instructions), followed by the Linux microVM with  (80k instructions).
The Nanos unikernel executes around 250k instructions on average during the same 1s idle period, which is 3 times more than the Linux microVM.
OSv executes considerably more instructions, at more than 9 million.
We initially suspected this to be a bug in the interaction between OSv and Firecracker, but additional measurements on QEMU show similar results, leading us to assume that this is specific to the OSv implementation.
\texttt{runc}-based containers do not execute enough instructions to consistently record data with \texttt{perf} as the containerized handler process is scheduled directly by the host kernel and thus blocked when no request is handled.
While \texttt{runc}-based containers have a decisive advantage here, the difference between gVisor and Linux microVMs is not as significant as expected, with the Linux microVM requiring less than twice as many instructions.
We were, however, surprised about the comparatively high idle CPU usage of unikernels.

\begin{figure*}
    \centering
    \begin{subfigure}{0.49\textwidth}
        \centering
        \includegraphics[width=0.975\columnwidth]{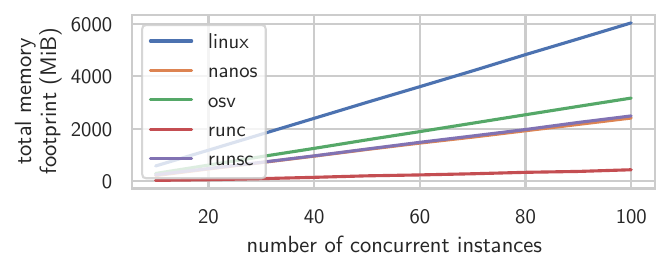}
        \caption{Go no-op function}
        \label{fig:idle-memory-all:go}
    \end{subfigure}%
    \hfill
    \begin{subfigure}{0.49\textwidth}
        \centering
        \includegraphics[width=0.975\columnwidth]{./graphs/plots/memory-go.pdf}
        \caption{Node.js no-op function}
        \label{fig:idle-memory-all:node}
    \end{subfigure}%
    \caption{Total memory usage for n instances of the same function}
    \label{fig:idle-memory-all}
\end{figure*}

\subsubsection*{Memory}
To measure the memory footprint of our sandbox mechanisms, we look at the \texttt{/proc/meminfo} interface on our host to determine the amount of available memory while scaling the number of concurrent idle function instances.
Having multiple idle instances of the same function is common in FaaS and should, in theory, enable benefits from kernel same-page merging.
Our results in \cref{fig:idle-memory-all} show a linear increase in memory consumption with increasing number of instances.
For the Go function, our results show that \texttt{runc}-based Docker containers have by far the lowest memory usage at only 3.9MiB per instance, which is probably the result of having parts of the function executable shared between containers (recall that the no-op function itself will not allocate significant amounts of memory).
When executing the same container with gVisor, memory usage is several times larger at 24.7MiB per instance.
This indicates an increased memory overhead from the gVisor sandbox, in addition to not being able to share the page cache for the function executable between containers.
The memory usage of Nanos and OSv is in the same order of magnitude as that of gVisor, while the Linux microVM required about twice the amount of memory.
The Node.js function exhibits a similar pattern, albeit with a higher overall memory usage.
The notable exception to this is Nanos, which requires 160.8MiB per instance for the Node.js function, which is the result of loading the entire Node.js executable into memory instead of relying on on-demand paging.

Our findings show that unikernels can require less than half the amount of per-instance memory compared to normal Linux microVMs and can reach levels similar to gVisor-based containers.
However, in its current implementation state, the lack of demand paging in Nanos increases memory usage in a way that makes it impractical for use with large binaries such as Node.js.
At the same time, we find that normal Docker containers are hard to beat in terms of memory usage due to their tight integration with the host system, which allows for a high degree of resource sharing.

\subsection{CPU Performance}
\label{sec:results:cpu}

\begin{figure}
    \centering
    \includegraphics[width=\linewidth]{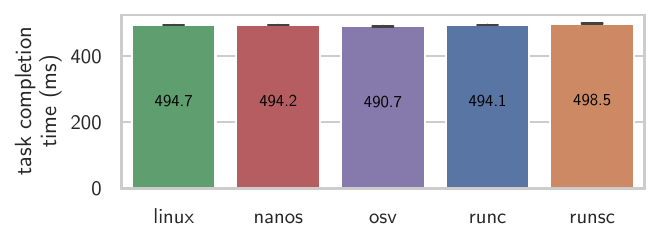}
    \caption{Mean time to calculate the $10^{9}$-th Fibonacci number}
    \label{fig:bench-fibonacci}
\end{figure}

We next consider CPU performance of our isolation mechanisms using a CPU-bound iterative calculation of the $10^{9}$th Fibonacci number implemented in Go.
We perform warm invocations for each operation.
As the calculation does not require file system access or other system calls all sandboxes perform equally well, as shown in \cref{fig:bench-fibonacci}

\subsection{Memory Performance}
\label{sec:results:memory}

\subsubsection*{Read}

\begin{figure}
    \centering
    \includegraphics[width=\linewidth]{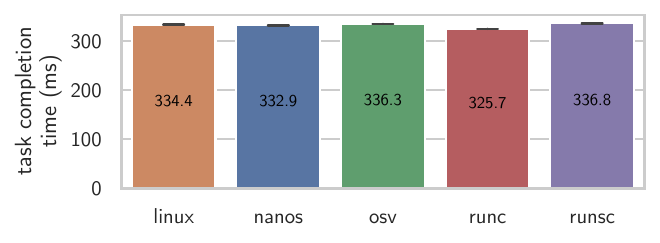}
    \caption{Mean time for a matrix multiplication of size $1024 \times 1024$}
    \label{fig:bench-matrix}
\end{figure}

To measure memory performance, we first deploy a matrix multiplication function written in Go to our testbed.
This function multiplies two $1024 \times 1024$ matrices, requiring extensive access to allocated memory pages, therefore not accounting for page fault costs.
The results in \cref{fig:bench-matrix} again show similar performance for all sandboxes.

\subsubsection*{Write}

\begin{figure}
    \centering
    \includegraphics[width=\linewidth]{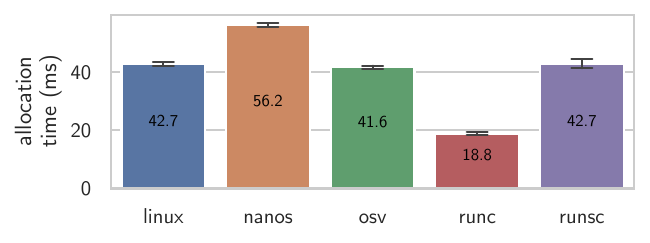}
    \caption{Mean time required to allocate 50MiB of memory}
    \label{fig:bench-alloc}
\end{figure}

We further evaluate the memory allocation performance by deploying a Go-based function that allocates 50MiB of contiguous virtual memory, which corresponds to 12,800 pages at a 4k page size.
To mitigate the effect of demand paging at both the operating system and hypervisor level, the function handler also writes a single byte to each allocated memory page to ensure that the allocation is backed by physical memory.

Our results in \cref{fig:bench-alloc} show similar latency values for Linux, OSv, and gVisor containers at 40ms on average, whereas Nanos takes 1.4 times longer.
Unsurprisingly, the allocation is more than twice as fast on containers with the \texttt{runc} runtime as allocation is performed directly by the host kernel, thus avoiding additional page faults at the hypervisor level.
gVisor implements its own memory management independent of the host kernel, which requires an additional layer of memory translation similar to that of a hypervisor when using gVisor's KVM backend.
Tracing of the \texttt{kvm\_mmu\_page\_fault} host kernel function with \texttt{bpftrace} shows between 13,100 and 14,400 page faults for this operation for Firecracker and gVisor (with the KVM backend), indicating that all allocations are based on newly allocated physical memory.
The longer allocation times in Nanos can therefore be directly attributed to its memory management implementation.

\subsection{Network I/O Performance}
\label{sec:results:network}

\subsubsection*{Concurrency}

\begin{figure}
    \centering
    \includegraphics[width=\linewidth]{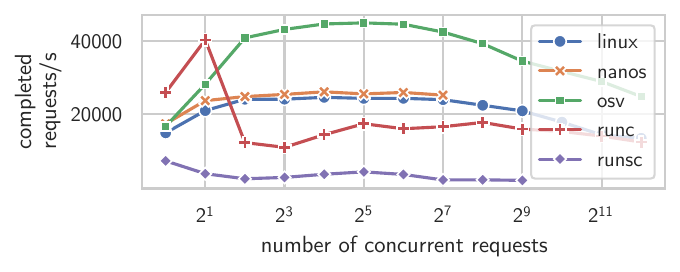}
    \caption{Requests per second for $2^{15}$ = 32,768 requests}
    \label{fig:network-rps}
\end{figure}

Where the trust model allows it, concurrently serving requests for the same function from a single function instance can boost both throughput and efficiency of a FaaS platform~\cite{paper_pfandzelter2020_tinyfaas,jia2021nightcore}.
To evaluate the capabilities of our sandboxes supporting concurrent function invocations, we again deploy the no-op function.
Instead of proxying through our experiment harness, we now connect our clients directly to function instances over HTTP (a highly scalable FaaS proxy is out of scope for this work).
Note that all requests are within a host and do not traverse physical networks.

\Cref{fig:network-rps} shows the result of using the \textit{hey} HTTP load generator\footnote{\url{https://github.com/rakyll/hey}} to issue $2^{15} = 32,768$ requests to this function.
We repeat this measurement with an increasing number of parallel client threads from 1 to 4,096 in power of two increments.
The results show a clear advantage for OSv, which is consistently able to handle more requests than all other environments with a maximum of 44,793 requests per second, whereas the Linux microVM handles a maximum of 24,621 requests per second.
Although Nanos performs slightly better for low numbers of concurrent requests, errors appear for more than 256 concurrent requests.
We expected \texttt{runc}-based containers to perform better than the Linux microVM, which they do for one and two concurrent requests.
However, we also observe a steep decline beyond two concurrent requests, where performance is lower than on the Linux microVM.
gVisor-based containers perform significantly worse than all other candidates and are unable to handle more than 512 concurrent requests.

The exceptional performance of OSv may to some extent be attributed to OSv's network stack being based on the FreeBSD implementation, which is known to be fast.
It has also been further optimized to use a design based on network channels to reduce locking~\cite{osv_2014}.
The network stack in Nanos, on the other hand, is based on the open-source \emph{lwIP} implementation~\cite{dunkels_design_2001}, which was originally created for use in embedded systems and will therefore hardly be able to compete with OSv or Linux.
gVisor implements its own network stack in the user-level Sentry process, which appears to come at the cost of considerably degraded performance.

\subsubsection*{Throughput}

\begin{figure}
    \centering
    \includegraphics[width=\linewidth]{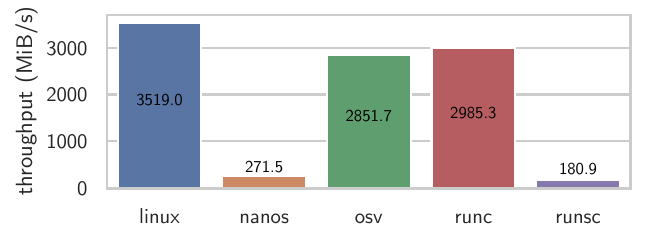}
    \caption{MiB/s network throughput}
    \label{fig:network-throughput}
\end{figure}

To measure network throughput, we deploy a Go-based function that serves a 50MiB static file.
We repeatedly call this function from four concurrent clients for a total transfer size of 10GiB.
The results in \cref{fig:network-throughput} show about 1.2 times higher throughput on the Linux microVM compared to OSv.
Curiously, OSv warns that it does not support the \texttt{sendfile} system call, likely used by the Go standard library HTTP server for serving static files, probably negatively impacting performance on OSv.
Nanos performs significantly worse, achieving a throughput of only 271MiB/s.
We observe slightly lower throughput on \texttt{runc}-based containers compared to the Linux microVM, which we did not expect.
gVisor-based containers again show the worst performance, reaching a throughput of around 180MiB/s.

\subsection{File System Performance}
\label{sec:results:filesystem}

\begin{figure}
    \centering
    \includegraphics[width=\linewidth]{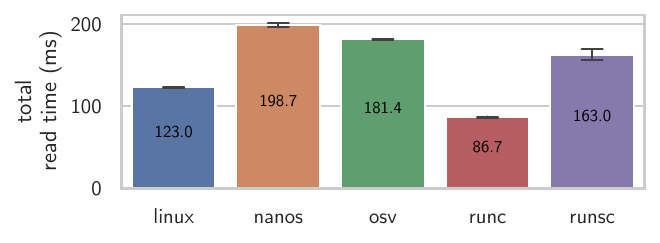}
    \caption{Time required to load a 50MiB file into memory}
    \label{fig:bench-read}
\end{figure}

Finally, we evaluate disk read performance by deploying a Go-based function that reads a static 50MiB file from disk and loads it into memory.
Note that disk read performance can be a factor in cold start performance for larger function handlers.
We did not evaluate disk \emph{write} performance as FaaS functions are unlikely to store new data locally.
We repeat this experiment 100 times, which likely gives containers and Firecracker the ability to store the function image or root file system in the host page cache.

The findings in \cref{fig:bench-read} show \texttt{runc} as the fastest sandbox (87ms on average).
Between Firecracker instances, the Linux microVM performs best at a mean 123ms.
Both Nanos and OSv take significantly longer, with around 199ms and 181ms, respectively.
OSv and Nanos have comparable read times, yet there is still a gap of about 60ms when compared to Linux.
This can be attributed to their file system or disk driver implementations, with all other factors being equal.
gVisor containers are slightly faster than OSv and Nanos, but still take around 1.3 times longer on average compared to the Linux microVM.
This demonstrates a clear advantage of the optimized disk driver and/or file system stack in Linux.

%% file: sections/5_discussion.tex
\section{Discussion \& Future Work}
\label{sec:discussion}

While our in-depth experimental evaluation reveals no clear `winner' for FaaS isolation, we have seen that unikernels should be regarded as viable alternatives, with Nanos and OSv outperforming Linux microVMs and gVisor in several metrics.
Nevertheless, we want to draw attention to some limitations of our study and discuss directions for future research.

\subsection{Common FaaS Optimizations}
\label{sec:discussion:optimization}

Commercial cloud FaaS systems use a number of optimizations to reduce the cold start overhead of functions, with many more proposed in academia~\cite{paper_schirmer2022_fusionize,paper_bermbach2020_faas_coldstarts,jia2021nightcore,schirmer2023profaastinate}.
For example, AWS Lambda removes the Linux microVM kernel boot time from a function's cold start hot path by using pre-booted virtual machines in which function handler code is then mounted~\cite{aws-lambda-internals}.
However, this only moves the overhead out of the critical path and increases complexity.
We found that not only do unikernels reduce boot times without additional configuration after starting, they consequently also require much less resources during this process.
Our results showed that compared to Nanos, Linux can require up to 8.4 times as many instructions to boot and start the Go function handler.
Combined with fast cold start times, this significantly improves the situation around cold starts compared to Linux microVMs, especially when considering constrained edge resources.
Further research on how common FaaS optimization techniques could be applied to unikernels is required, yet outside the scope of this paper.
For example, restoring (Linux and unikernel) microVM snapshots to reduce the cold start time is possible, albeit not a zero-cost operation~\cite{faasnap, serverless-function-snapshots}.

\subsection{Inefficient Function Runtimes}

Although we found that the benefits of unikernels hold even for more heavy-weight functions, such as our Node.js function, we note that the relative advantages become much smaller.
Our experiments showed that the overhead of starting even our very simple Node.js test function can easily exceed that of booting a unikernel multiple times and similarly increase cold start times.
We infer from this that choosing an efficient programming language to implement FaaS workloads is at least as important as the sandboxing technology used.
Statically compiled programming languages such as Go or Rust can contribute considerably to the overall efficiency of the FaaS system and help to keep cold start times short.
Unikernels can improve cold start efficiency and memory footprint, but cannot compensate for the inefficiencies introduced by the user-provided function code.

\subsection{Efficiency of Containers}

While we only included \texttt{runc} containers for reference given their prevalence in open-source FaaS systems with arguably lower security requirements, our results clearly showed their performance benefits.
The tight integration with the host operating system allows for effective resource sharing, e.g., giving containers access to the host page cache.
We further saw that a \emph{secure} container runtime such as \texttt{runsc} trades most of these performance benefits for security.
As such, future research should investigate making this trade-off configurable, e.g., allowing functions from the same tenant or for the same client to exist in containers while using virtualization only where necessary to guarantee isolation.

\subsection{Unikernel Choice}

We chose Nanos and OSv as mature, Linux-compatible examples of the unikernel and saw that performance characteristics between the two can vary in either direction.
For a more universal claim on whether unikernels are a viable sandbox alternative for edge FaaS, evaluating further unikernels or even designing a custom unikernel for this use case is necessary.
For example, we had originally considered the \emph{Unikraft}~\cite{kuenzer_unleashing_2019} unikernel, but found it lacking both sufficient Firecracker and Linux compatibility.

\subsection{Language-based Sandboxes}

We only considered Linux-compatible sandbox mechanisms for our evaluation, as we found them the most promising for widespread adoption in a FaaS system.
Nevertheless, several language-specific unikernels exist, e.g., Clive for Go~\cite{clive}, runtime.js for JavaScript~\cite{runtimejs}, or Hermit for Rust~\cite{lankes_rustyhermit_2020}.
Through optimization for a specific language, they might provide better performance for some workloads.
In fact, as FaaS systems usually require users to specify their runtime and functions are small, it may be possible to select a specific unikernel for each function.

\subsection{Usability}

Familiar programming abstractions are a key selling point of the FaaS paradigm.
It is certainly possible to achieve high levels of Linux compatibility, as demonstrated by the production use of gVisor or other Linux compatibility layers, such as in FreeBSD~\cite{freebsd-linuxulator}.
However, these will never achieve perfect compatibility, and there will always remain edge cases that can break existing applications, however unlikely in small FaaS workloads.
Further, debugging problems on a unikernel is challenging due to their single-process nature and the complete lack of onboard debugging facilities~\cite{cantrill-rant}, which requires attaching debuggers to the VMM and debugging both the kernel and the application at the same time.

\subsection{VMM Complexity}

Although Firecracker is already a highly minimal VMM, the requirement to run a full Linux kernel requires implementing a complex interface.
The adoption of unikernels could further reduce this interface and its complexity, as suggested by Williams and Koller~\cite{ukvm_2016}, in turn also improving security.
For example, the \emph{solo5} unikernel middleware~\cite{solo5github} has been developed with an approach that incorporates a minimal set of hyper calls.
These hyper calls enable unikernel guests to send and receive network packets, interact with block devices to create file systems, and access the system time.
Moving towards such simple abstractions will be crucial to ensure both efficiency and security for unikernels.

%% file: sections/6_relwork.tex
\section{Related Work}
\label{sec:relwork}

We are not the first to explore unikernels as an isolation mechanism in FaaS platforms:
Early work on unikernels by Koller and Williams~\cite{ibm_2017} has outlined their potential benefits for serverless workloads and questioned whether Linux will be able to provide the necessary lightweight sandboxed execution environments.
G\'{e}hberger and Kov\'{a}cs~\cite{cooling-down-faas} and Mistry et al.~\cite{faas-unikernel-practical} presented prototypical FaaS platforms built around IncludeOS (C++) and MirageOS (OCaml) language-specific unikernels.
In both cases, the authors found that unikernels offer a viable alternative to container-based approaches, particularly by reducing cold start times and memory usage, yet it is unlikely that language-specific unikernels will be widely adopted.

Goethals et al.~\cite{goethals_2022} compared OSv unikernels, \texttt{runc} and \texttt{runsc}-based containers, and Firecracker paired with Linux microVMs for use in FaaS deployments.
Their work agrees with our findings, yet does not provide an in-depth cost comparison, especially in terms of the overhead incurred when instantiating the execution environment and during idle periods.
Our work also assesses the impact of different programming languages to implement functions, which we found to be an important factor.
Finally, our evaluation included Nanos as an additional unikernel, showing significant performance difference between unikernel implementations.

WebAssembly implements a secure abstract machine based on a linear memory model, which provides a light-weight isolation mechanism.
This has led to an extensive body of literature focused on employing it for serverless and FaaS use cases~\cite{hall_2019,gadepalli_2020, gackstatter_2022}.
Work in this area largely finds WebAssembly promising yet notes challenges in performance and application compatibility.
Further, the isolation and security provided by the WebAssembly sandbox alone will be insufficient for multitenancy~\cite{security_wasm,swivel_wasm}.
This is exacerbated when removing additional security boundaries afforded by the traditional process model, e.g., \texttt{seccomp} profiles, to increase performance~\cite{gadepalli_2020}.

%% file: sections/7_conclusions.tex
\section{Conclusion}
\label{sec:conclusion}

FaaS is a promising programming and execution paradigm for edge applications, but the question of isolating serverless functions at the edge efficiently has not been answered conclusively.
In this paper, we proposed the idea of unikernels as a secure edge FaaS execution environment.
In extensive evaluation of the Linux-compatible Nanos and OSv unikernel tool chains, we demonstrated advantages in specific aspects such as cold start efficiency and memory usage.
In their current state, however, they do not offer the same level of stability and reliability as Linux and require more technical expertise to be used successfully.
We believe that unikernels can offer a viable alternative to existing approaches in the future -- if they receive sufficient investment in their advancement.